\documentclass[10pt,conference,compsocconf]{IEEEtran}

\usepackage{cite,url,color}
\usepackage{graphicx} 
\graphicspath{{./pdf/}{./jpeg/}{./figs/}{./eps/}}
\DeclareGraphicsExtensions{.pdf,.jpeg,.png,.eps}
\usepackage{multirow}
\usepackage[cmex10]{amsmath}
\usepackage{amssymb,bm}
\let\OLDthebibliography\thebibliography
\renewcommand\thebibliography[1]{
  \OLDthebibliography{#1}
  \setlength{\parskip}{0pt}
  \setlength{\itemsep}{0pt}
}

\usepackage{algpseudocode} 
\newcommand{\algrule}[1][.5pt]{\par\vskip.25\baselineskip\hrule height #1\par\vskip.25\baselineskip}
\usepackage[caption=false,font=footnotesize]{subfig}
\usepackage[amsmath,thmmarks,hyperref]{ntheorem}

\newcommand{\ve}[1]{\mathbf{{#1}}}
\newcommand{\abs}[1]{\ensuremath{\vert #1\vert}}
\newcommand{\norm}[1]{\ensuremath{\Vert #1\Vert}}

\begin{document}

\title{Compression via Compressive Sensing : A Low-Power Framework for the Telemonitoring of Multi-Channel Physiological Signals}%

\author{%
\IEEEauthorblockN{Benyuan~Liu}%
\IEEEauthorblockA{%
ATR Laboratory, \\
National University of Defense Technology\\
Changsha, Hunan, 410074, P. R. China\\
E-mail: liubenyuan@gmail.com}%
\and
\IEEEauthorblockN{Zhilin~Zhang}%
\IEEEauthorblockA{%
Samsung Research America - Dallas\\%
Richardson, TX 75082, USA\\%
E-mail: zhilinzhang@ieee.org}%
\and
\IEEEauthorblockN{Hongqi~Fan,~Qiang~Fu}%
\IEEEauthorblockA{%
ATR Laboratory, \\
National University of Defense Technology\\
Changsha, Hunan, 410074, P. R. China\\
E-mail: fanhongqi@tsinghua.org}%
}%

\maketitle

\begin{abstract}
    Telehealth and wearable equipment can deliver personal healthcare and necessary treatment remotely. One major challenge is transmitting large amount of biosignals through wireless networks. The limited battery life calls for low-power data compressors. Compressive Sensing (CS) has proved to be a low-power compressor. In this study, we apply CS on the compression of multichannel biosignals. We firstly develop an efficient CS algorithm from the Block Sparse Bayesian Learning (BSBL) framework. It is based on a combination of the block sparse model and multiple measurement vector model. Experiments on real-life Fetal ECGs showed that the proposed algorithm has high fidelity and efficiency. Implemented in hardware, the proposed algorithm was compared to a Discrete Wavelet Transform (DWT) based algorithm, verifying the proposed one has low power consumption and occupies less computational resources.
\end{abstract}


\section{Introduction}
Personal healthcare benefits from the development of wireless telemonitoring. Using on-body sensors of these telemonitoring systems, biosignals reflecting health status can be easily collected and transmitted, and diagnosis is thus performed remotely at a data central. This client--central model calls for low-power telemonitoring techniques\cite{duarte2012signal}. One major challenge is huge amount of data collected versus limited battery life of portable devices and limited bandwidth of wireless networks. Signals need to be compressed before transmitting. However, most traditional compression methods consist of sophisticated matrix-vector multiplication and encoding which subsequently drains the battery.

Compressive sensing (CS) can be used as an efficient, lossy compression method. In this framework, the complexity is shifting from battery driven devices to data centrals, where a CS algorithm is used to recover original recordings. In practice, telemonitoring systems allow users to move freely or engage in activities. Artifacts caused by body movements deteriorate the quality of signals\cite{Zhang_Asilomar,martin2000issues} and may result in varied sparsity. These pose challenges for CS algorithms to fidelity recover the biosignals. Recently, Zhang et al. \cite{Zhang2012a} proposed the Block Sparse Bayesian Learning (BSBL) framework, which showed excellent performance when recovering less-sparse signals with high fidelity. However, to recovery multichannel signals, existing BSBL algorithms \cite{Zhang2012a,liu2013energy} have to recover them channel-by-channel.

Recently a spatio-temporal sparse Bayesian learning model was proposed for compressive sensing of multichannel biosignals \cite{Zhang_Asilomar}, which is a combination of the block sparse model \cite{Zhang2012a} and the multiple measurement vector model \cite{Zhang2011}. In this work we extend our fast BSBL algorithm \cite{liu2013energy} for this model, such that it is suitable for compressed sensing of multi-channel biosignals with fast speed. It has high fidelity and efficiency in recovering real-life fetal ECGs. We also provide the evaluation of CS-based and DWT-based compressors in the platform of Field Programmable Gate Array (FPGA). Results show that the CS-based framework has fewer compression latency, consumes less resource and dynamic power.

Throughout the paper, {\bf bold} letters are reserved for vectors and matrices. The concatenation of scalars and vectors are denoted by $\ve{y}=\{y_1,\ldots,y_S\}$ and $\ve{X}=\{\ve{x}_1,\ldots,\ve{x}_N\}$. $\otimes$ denotes the Kronecker product and $\mathrm{Tr}(\ve{A})$ denotes the trace of $\ve{A}$. $\mathrm{vec}(\ve{A})$ represents the vectorization of $\ve{A}$. $\ve{I}_d$ denotes an identity matrix with size $d$.

\section{The Proposed Algorithm}
The Multiple-Measurement Vector (MMV) model\cite{Zhang2011} can be applied to multi-channel physiological signals $\ve{X}$, where $\ve{X}\triangleq\{\ve{x}^1,\cdots,\ve{x}^P\}$ and $\ve{x}^i\in\mathbb{R}^{N\times 1}$ is a column vector represents the data samples of the $i$th channel. $\ve{X}\in\mathbb{R}^{N\times P}$ is also called a \emph{packet}. The compressed measurements $\ve{Y}$ is obtained as
\begin{equation}
    \ve{Y} = \bm{\Phi}\ve{X},
\end{equation}
where $\bm{\Phi}$ is the sensing matrix. To minimize the power consumption, Bernoulli sensing matrix $\bm{\Phi}\in\{0,1\}^{M\times N}$ whose entries consist of $0$s and $1$s is preferred \cite{Zhang_TBME2012b,liu2013energy}.

\begin{figure}[!ht]
\centerline{\includegraphics[width=1.5in]{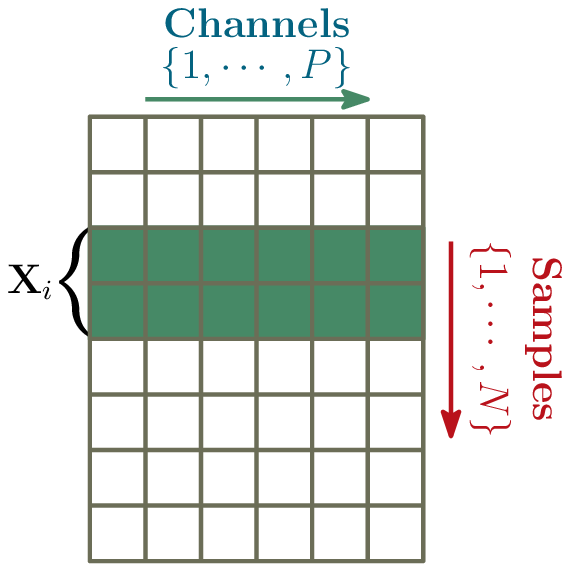}}
\caption{The illustration of a \emph{packet} and a Block MMV block $\ve{X}_i$ in multi-channel physiological signals.}
\label{fig:bsbl_mmv}
\end{figure}

As in \cite{Zhang_Asilomar}, the block sparse structure \cite{Zhang2012a} is incorporated into the above MMV model. In Fig \ref{fig:bsbl_mmv}, the $i$th (where $i\in\{1,\cdots,g\}$) block is denoted by $\ve{X}_i$, which has the size $d_i\times P$.
A packet $\ve{X}$ is therefore divided into $g$ blocks. We then extend our fast BSBL algorithm \cite{liu2013energy} for this model.

The prior of the $i$th block $\ve{X}_i$ is modeled using the matrix-variate Gaussian distribution
\begin{equation}\label{eq:xi}
p(\ve{X}_i; \gamma_i) = \mathcal{MN}(\ve{X}_i;\ve{0},\gamma_i\ve{I}_{d_i},\ve{I}_P)
\end{equation}
which can be re-written as
\begin{equation}
    p(\mathrm{vec}(\ve{X}_i); \gamma_i) = \mathcal{N}(\mathrm{vec}(\ve{X}_i);\ve{0}, \gamma_i\ve{I}_{d_i\cdot P})
\end{equation}
where $\gamma_i$ is a variance parameter, and $\gamma_i\ve{I}_{d_i\cdot P}=\ve{I}_P\otimes\gamma_i\ve{I}_{d_i}$. The learning process of $\gamma_i$ automatically determines the relevance (whether it is a zero block or not) of the $i$th block\cite{Zhang_Asilomar}. Note that here we explicitly ignore the intra-block correlation\cite{Zhang2012a} as well as the correlation among channels\cite{Zhang2011} for fast implementation. We further assume that blocks are mutually uncorrelated and write the prior of a packet $\ve{X}$ as,
\begin{equation}\label{eq:x_model}
p(\ve{X}; \{\gamma_i\}) = \mathcal{MN}(\ve{X};\ve{0},\bm{\Gamma},\ve{I}_P)
\end{equation}
where $\bm{\Gamma}$ is a block diagonal matrix with the $i$th principle diagonal given by $\gamma_i\ve{I}_{d_i}$.

The measurements $\ve{Y}$ is modeled as
\begin{equation}\label{eq:y_model}
p(\ve{Y}|\ve{X};\beta) = \mathcal{MN}(\ve{Y}|\bm{\Phi}\ve{X};\beta^{-1}\ve{I}_M,\ve{I}_P)
\end{equation}
where measurement noise is assumed to be i.i.d Gaussian with the precision parameter given by $\beta$.

Given \eqref{eq:x_model} and \eqref{eq:y_model}, we derive the posterior $p(\ve{X}|\ve{Y}; \{\gamma_i\}, \beta)$ as follows
\begin{align}
p(\ve{X}|\ve{Y};\{\gamma_i\},\beta) &= \mathcal{MN}(\ve{X};\bm{\mu},\bm{\Sigma},\ve{I}_P)
\end{align}
where
\begin{align}
\bm{\mu} &= \beta\bm{\Sigma}\bm{\Phi}^T\ve{Y} \\
\bm{\Sigma} &= (\bm{\Gamma}^{-1} + \beta\bm{\Phi}^T\bm{\Phi})^{-1}. \label{eq:c0}
\end{align}

Similar as in \cite{liu2013energy}, the Type II Maximum Likelihood method  is used to estimate the parameters $\{\gamma_i\},\beta$, which leads to the following cost function,
\begin{align}
\mathcal{L}(\{\gamma_i\},\beta) &\triangleq -2\log p(\ve{Y};\{\gamma_i\},\beta) \\
 &= N\log \abs{\ve{C}} + \mathrm{Tr}\left[ \ve{Y}^T\ve{C}^{-1}\ve{Y} \right] \label{eq:loglikelihood}
\end{align}
where  $\ve{C} = \beta^{-1}\ve{I}_M + \bm{\Phi}\bm{\Gamma}\bm{\Phi}^T$.


We optimize the cost function \eqref{eq:loglikelihood} with Fast Marginalized Likelihood Maximization (FMLM) method\cite{liu2013energy}. Let $\bm{\Phi}_i\in \mathbb{R}^{M\times d_i}$ be the $i$th column block in $\bm{\Phi}$, $\ve{C}$ in \eqref{eq:c0} can be rewritten as:
\begin{align}
    \ve{C} &= \beta^{-1}\ve{I} + \sum_{m\neq i}^{g}
\bm{\Phi}_m\gamma_m\bm{\Phi}_m^T+
\bm{\Phi}_i\gamma_i\bm{\Phi}_i^T \\
 &= \ve{C}_{-i} + \bm{\Phi}_i\gamma_i\bm{\Phi}_i^T \label{eq:c}
\end{align}
where $\ve{C}_{-i} \triangleq \beta^{-1}\ve{I} + \sum_{m\neq i}
\bm{\Phi}_m\gamma_m\bm{\Phi}_m^T$.
Using the Woodbury Identity, \eqref{eq:loglikelihood} can be rewritten as:
\begin{align}
\mathcal{L} =& N\log\abs{\ve{C}_{-i}} + \mathrm{Tr}\left[\ve{Y}^T\ve{C}_{-i}^{-1}\ve{Y}\right]
\nonumber \\
 &+ N\log\abs{\ve{I}_{d_i} + \gamma_i\ve{s}_i} - \mathrm{Tr}\left[\ve{q}_i^T(\gamma_i^{-1}\ve{I}_{d_i} + \ve{s}_i)^{-1}\ve{q}_i\right]
\nonumber \\
 =& \mathcal{L}(-i) + \mathcal{L}(i) \nonumber
\end{align}
where
$\ve{s}_i\triangleq \bm{\Phi}_i^T\ve{C}_{-i}^{-1}\bm{\Phi}_i$,
$\ve{q}_i\triangleq\bm{\Phi}_i^T\ve{C}_{-i}^{-1}\ve{Y}$ and
\begin{equation*}
\mathcal{L}(i) \triangleq N\log\abs{\ve{I}_{d_i} + \gamma_i\ve{s}_i} -
\mathrm{Tr}\left[ \ve{q}_i^T(\gamma_i^{-1}\ve{I}_{d_i} + \ve{s}_i)^{-1}\ve{q}_i\right]
\end{equation*}
$\gamma_i$ can be efficiently updated by optimizing over $\mathcal{L}(i)$,
\begin{equation}\label{eq:gamma_0}
\gamma_i =
\frac{1}{d_i}\mathrm{Tr}\left[\ve{s}_i^{-1}(\ve{q}_i\ve{q}_i^T - \ve{s}_i)\ve{s}_i^{-1}\right].
\end{equation}

The proposed algorithm (denoted as \textbf{MBSBL-FM}) is given in Fig. \ref{algo:bsbl-fm}.
\begin{figure}[h!]
\centering
\begin{algorithmic}[1]
    \algrule
\Procedure{MBSBL-FM}{$\ve{Y}$,$\bm{\Phi}$,$\eta$}
\State Outputs: $\ve{X},\bm{\Sigma},\bm{\gamma}$
\State Initialize $\beta^{-1}= 0.01\norm{\ve{Y}}_\mathcal{F}^2$
\State Calculate $\{\ve{s}_i\}$, $\{\ve{q}_i\}$
\While{not converged}
\State Calculate $\tilde{\gamma_i} = \frac{1}{d_i}\mathrm{Tr}\left[\ve{s}_i^{-1}(\ve{q}_i\ve{q}_i^T - \ve{s}_i)\ve{s}_i^{-1}\right], \forall i$
\State Calculate $\Delta \mathcal{L}(i) = \mathcal{L}(\tilde{\gamma_i}) - \mathcal{L}(\gamma_i), \forall i$
\State Select the $\hat{i}$th block s.t. $\Delta\mathcal{L}(\hat{i})=\min\{\Delta\mathcal{L}(i)\}$
\State Re-calculate $\bm{\mu},\bm{\Sigma},\{\ve{s}_i\},\{\ve{q}_i\}$
\State Re-calculate the convergence criterion
\EndWhile
\EndProcedure
    \algrule
\end{algorithmic}
\caption{The MBSBL-FM Algorithm.}
\label{algo:bsbl-fm}
\end{figure}
Within each iteration, it only updates the most relevant block that attributes to the deepest descent of \eqref{eq:loglikelihood}. The detailed procedures on re-calculation of $\bm{\mu},\bm{\Sigma},\{\ve{s}_i\},\{\ve{q}_i\}$ are similar to \cite{liu2013energy}. The algorithm terminates when the maximum change of the cost function is smaller than the threshold $\eta$.

{\flushleft\bf Remark I:}
$\beta$ can be estimated from \eqref{eq:loglikelihood}. However, in our wireless telemonitoring applications, the measurement noise $\mathbf{V}$ can be ignored (artifacts in biosignals are incorporated into $\mathbf{X}$). Thus, one can set $\beta^{-1}$ to a very small value.  Our experiments showed that  $\beta^{-1}=0.01\norm{\ve{Y}}_\mathcal{F}^2$ led to satisfactory results. Besides, we set the threshold $\eta=1\mathrm{e}^{-5}$.

{\flushleft\bf Remark II:}
Many biosignals (especially those recorded during ambulatory telemonitoring) are not sparse directly in the time domain \cite{Zhang_Asilomar,martin2000issues}, therefore people often resort to a transformed domain in firstly seeking the sparse coefficients $\ve{A}$, which can be expressed as
\begin{equation}\label{eq:cs_in_d}
\ve{Y} = (\bm{\Phi}\ve{D}) \ve{A},
\end{equation}
where $\ve{D}$ may be a Discrete Cosine Transform (DCT) Matrix and $\ve{A}\triangleq\{\bm{\alpha}_1,\ldots,\bm{\alpha}_P\}$ were the transformed coefficients, i.e., $\mathbf{x}^i = \mathbf{D} \bm{\alpha}_i$. Then we obtain the reconstructed signal $\hat{\ve{X}}$ by $\hat{\ve{X}} = \ve{D}\ve{A}$.

\section{Telemonitoring of Multi-Channel Fetal ECGs via Compressive Sensing}
We conducted experiments using real-life Fetal ECGs\cite{Zhang_TBME2012b}. The recordings contained  abdominal 8-channel signals sampled at $256$Hz. This dataset is hard for most existing CS algorithms to decompress. Only BSBL algorithms\cite{Zhang_TBME2012b,liu2013energy} showed high fidelity. In this experiment, we compared $11$ state-of-the-art CS solvers as shown in Table \ref{tab:cs}. The recovery was performed in the DCT transformed domain using \eqref{eq:cs_in_d}.
\begin{table}[!ht]
\renewcommand{\arraystretch}{1.3}
\centering
\caption{The CS recovery algorithms used in this paper.
}%
\label{tab:cs}
\begin{tabular}{lcc}
\hline\hline
{\bf CS Algorithm} & {\bf Objective Function} & {\bf Types}$^*$ \\
\hline
BP\cite{Berg2008}           & $\ell_1$ minimization & \\
SL0\cite{Mohimani2008}      & smooth $\ell_0$ minimization & \\
\hline
Group BP\cite{Berg2008}     & group $\ell_1$ minimization & \textcircled{1} \\
BSBL-EM\cite{Zhang2012a}    & Block Sparse Bayesian Learning & \textcircled{1}\textcircled{3} \\
BSBL-BO\cite{Zhang2012a}    & Block Sparse Bayesian Learning & \textcircled{1}\textcircled{3} \\
\textbf{BSBL-FM}\cite{liu2013energy} & \textbf{Block Sparse Bayesian Learning} & \textbf{\textcircled{1}} \\
\hline
SPG-MMV\cite{Berg2008}      & MMV $\ell_1$ minimization & \textcircled{2} \\
ISL0\cite{Hyder2010}        & MMV smooth $\ell_0$ minimization & \textcircled{2} \\
T-MSBL\cite{Zhang2011}      & MMV Sparse Bayesian Learning & \textcircled{2}\textcircled{4} \\
\hline
\textbf{MBSBL-FM}           & \textbf{MMV Block Sparse Bayesian Learning} & \textbf{\textcircled{1}\textcircled{2}} \\
\hline\hline
\multicolumn{2}{l}{$^*$\textcircled{1}: Block Sparse Model.} \\
\multicolumn{2}{l}{$^*$\textcircled{2}: MMV Model.} \\
\multicolumn{2}{l}{$^*$\textcircled{3}: Intra-block Correlation information.} \\
\multicolumn{2}{l}{$^*$\textcircled{4}: Inter-Vector Correlation information.} \\
\end{tabular}
\end{table}

The dataset was divided into packets. Each packet $\ve{X}\in\mathbb{R}^{N\times P}$ was constructed with $1$s ECG recordings where $N=256$ and $P=8$. We ran the experiment for 10 trials. In each trial, a new Bernoulli sensing matrix $\bm{\Phi}\in\{0,1\}^{M\times N}$ with exactly two $1$s of each column was generated. The compression ratio $\mathrm{CR}=(N-M)/N$ was varied from $0.4$ to $0.8$. With each parameter setting, the Normalized Mean Square Error ($\mathrm{NMSE}\triangleq\norm{\ve{\hat{X}} - \ve{X}}_\mathcal{F}^2/\norm{\ve{X}}_\mathcal{F}^2$) and  CPU time were recorded. The computer used in the experiments had a 2.9GHz CPU and 16GB RAM.

\begin{figure}[!htb]
    \centering
    \includegraphics[width=3.4in]{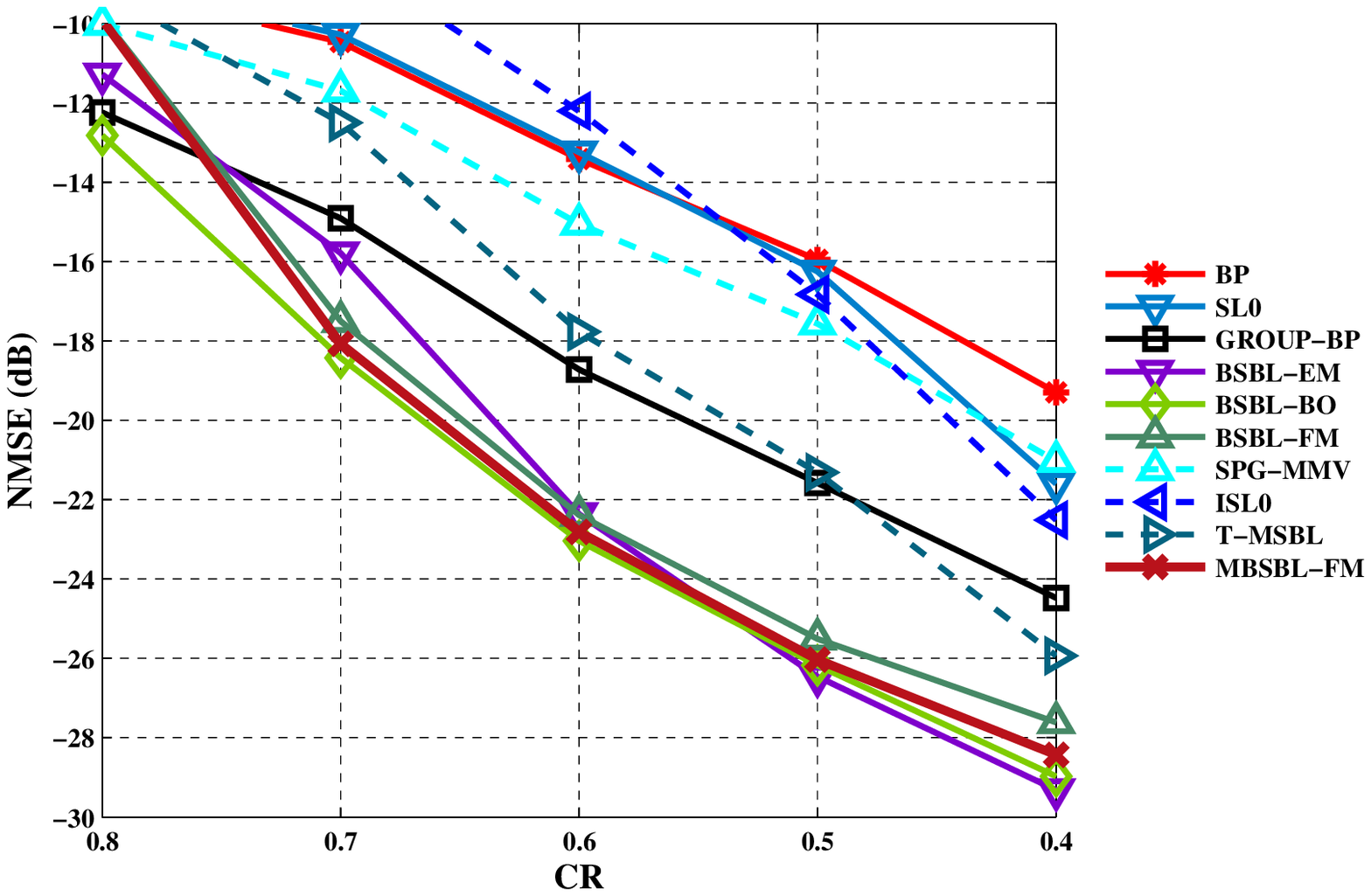} \\
    \includegraphics[width=3.28in]{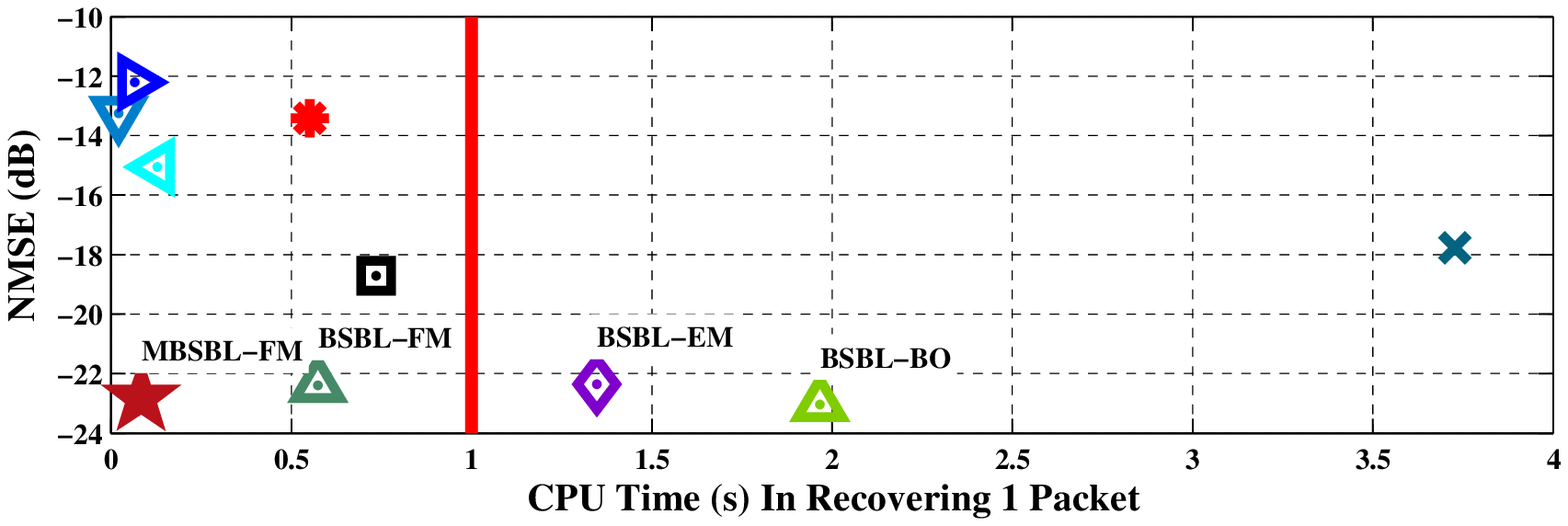}
    \caption{The average NMSE of recovering the FECG dataset with different compression ratios (Top) and the CPU runtime in recovering $1$ packet (Bottom, CR=$0.60$). The solid line in the plot indicates the 1-second duration of a packet.}
    \label{fig:exp_cr}
\end{figure}

From Fig \ref{fig:exp_cr}, we clearly see that
only BSBL algorithms showed the best performance with CR ranged from $0.4$ to $0.7$. The proposed algorithm, MBSBL-FM, was almost $24$ times faster than BSBL-BO and $17$ times faster than BSBL-EM, while still yielded similar NMSE value. It achieved a good balance between speed and performance.

\section{Hardware Evaluation}
In this section the energy consumption of DWT-based and CS-based compressors was evaluated on Field-Programmable Gate Array (FPGA). Unlike DSP or embedded platforms \cite{mamaghanian2011compressed}, FPGA favors parallel implementation and fix-point arithmetic. It implements only the logics related to the compressor while the rest are holding reset. Therefore, it is more suitable for low-energy applications.

The FPGA uses Registers (or Flip-Flops, denoted as FF) to store the bit-information and Look-Up-Tables (denoted as LUT) to implement the combinational logics. Data are stored in on-chip Memories (denoted as RAM). The dynamic power consumption (denoted as $P_d$) of a compressor reflects the design activity and switching events in the chip. In order to achieve low power, the chip should minimize circuitry activities and avoid using multipliers (denoted by MUL).

\subsection{Implementations}
The DWT-based compressor implemented in \cite{liu2013energy} was used for comparison. It adopted lift-based filtering and used the multiplierless LeGall 5/3 wavelet filter. The transformation stages of the DWT was set to $4$.

For each single-channel biosignal, the CS-based compressor consists only one matrix-vector multiplication, i.e., $\ve{y}=\bm{\Phi}\ve{x}$. In FPGA, the compression can be fully-paralleled with the \emph{on-the-fly} scheme: Let $\bm{\phi}_i$ denotes the $i$th column of the sensing matrix $\bm{\Phi}$, $\ve{y}^{(k)}$ denotes the compressed data vector after have received $k$ samples $\{x_1,\cdots,x_k\}$. Starts with $\ve{y}^{(0)}=\ve{0}$, the compressed data vector is iteratively updated with each new sample $x_i$,
\begin{equation}\label{eq:on-the-fly}
\ve{y}^{(i)} = \bm{\phi}_i x_i + \ve{y}^{(i-1)},
\end{equation}
The compression is done once the last sample $x_N$ of a packet has been acquired. When $\bm{\Phi}$ is the Gaussian sensing matrix, each $\bm{\phi}_i$ is real valued. \eqref{eq:on-the-fly} must be calculated in $M$ clock cycles with one multiplier unit. However, when $\bm{\Phi}$ is the Bernoulli matrix with two $1$s of each column, \eqref{eq:on-the-fly} can be implemented in one clock-cycle without multiplier. This is the reason why using the Bernoulli matrix can reduce energy consumption and save other hardware resources.

Note that the above procedure is performed  on all biosignals from different channels in a parallel way.

\subsection{Evaluation}

Table \ref{tab:fpga} shows the power consumption and other used hardware resources by the DWT-based compressor and the CS-based compressor when compressing a single-channel biosignal \footnote{For clarity, we only present the statistics on a single-channel biosignal. For multichannel biosignals, the advantages are more significant.}. The CS-based compressor adopted two kinds of sensing matrices. One was the random Gaussian sensing matrix (denoted by CS-Gaussian). The second was the Bernoulli sensing matrix as described above (denoted by CS-Bernoulli). In addition, we calculated the compression latency (denoted by $t_l$) which is the number of the clock cycles used for compressing after a packet has been acquired.
\begin{table}[!ht]%
\renewcommand{\arraystretch}{1.3}%
\centering%
\caption{The Compression Latency, Resources and Power Consumption of the DWT-based and the CS-based Compressors When Compressing a Single-channel Biosignal}%
\label{tab:fpga}%
\begin{tabular}{c|c|cccc|c}
\hline\hline
& $t_l$ & FF & LUT & RAM & MUL &  $P_d$ \\
\hline
DWT & 487   & 217 & 290  & 4     & 0 & 15mW \\
CS-Gaussian  & 387   & 80  & 96   & 10    & 1 & 21mW \\
CS-Bernoulli &\textbf{2}&\textbf{63}&\textbf{90}&\textbf{3}& 0 & \textbf{10}mW \\
\hline\hline
\end{tabular}
\end{table}

The CS-Gaussian compressor required one multiplier and multiple clock-cycles to generate the compressed measurements. It also needed   to store the whole quantized sensing matrix. Thus this implementation consumed more RAMs and energy. On the other hand, both the DWT-based compressor and the CS-Bernoulli compressor were multiplierless. The CS-Bernoulli had minimal latency, consumed only $\sim1/3$ Registers and LUTs than the DWT-based one. In order to reduce the RAM usage, it stored the locations of $1$s in the sensing matrix. Totally, the CS-Bernoulli compressor saved $31.25$\% dynamic power  against the DWT implementation.

\section{Conclusion}
In this paper, we proposed a fast compressive sensing algorithm that is based on a hybrid model of the block sparse model and the MMV model. It is suitable for compressive sensing of multichannel biosignals. Experiments on real-life Fetal ECGs showed that the proposed algorithm has both high fidelity and efficiency. We also compared the consumed hardware resources by the proposed algorithm and a DWT-based compression algorithm during data compression. The results showed that the proposed algorithm (adopting the Bernoulli sensing matrix) required shorter compression latency, less computational resources, and less power consumption.

\bibliographystyle{IEEEtran}
\bibliography{bsbl}

\end{document}